Short Paper*

# Towards the Development of 3D Engine Assembly Simulation Learning Module for Senior High School

John Paul P. Miranda
Mexico Campus, Don Honorio Ventura State University
jppmiranda@dhvsu.edu.ph
(corresponding author)

Jaymark A. Yambao
Mexico Campus, Don Honorio Ventura State University
jayambao@dhvsu.edu.ph

Jhon Asley M. Marcelo
Mexico Campus, Don Honorio Ventura State University
marcelojhonasley@gmail.com

Christopher Robert N. Gonzales
Mexico Campus, Don Honorio Ventura State University
chrisrg367@gmail.com

Vee-jay T. Mungcal
Mexico Campus, Don Honorio Ventura State University
mungcalveejay@gmail.com

Robine J. Baluyut
Mexico Campus, Don Honorio Ventura State University
binebinebaluyut20@gmail.com

**Abstract**

*Purpose* – The focus of the study is to develop a 3D engine assembly simulation learning module to address the lack of equipment in one senior high school in the Philippines.

*Method* – The study used mixed-method to determine the considerations needed in developing an application for educational use particularly among laboratory/practical subjects like engine assembly. The study used ISO 25010 quality standards in evaluating the application (*n* = 153 students and 3 ICT experts).

*Results* – Results showed that the application is moderately acceptable (overall mean = 3.52) under ISO 25010 quality standards.

*Conclusion* – The study created an engine simulation learning assembly in which teachers can use to augment their lesson. The study also highlights the applicability of using 3D-related technologies for practical and laboratory subjects particularly highly technical-related subjects.

*Recommendations* – Future studies may develop a similar application in the same context using mobile and other emerging technology (i.e., Virtual Reality, Augmented Reality) as well as making the content more customizable. Effectivity of the system in an actual setting is also worth pursuing.

*Research Implications* – The study highlighted the potential use of 3D technology in a classroom setting.

## INTRODUCTION

The simulation program is used to facilitate students in their learning by mimicking the actual scenario or situation. This allows the student to work, learn, experience even without the supervision of their professor or teacher (So, Chen, & Wan, 2019; Lateef, 2010). In the recent event, where the world is experiencing a health crisis due to the COVID-19 pandemic, the education sector is looking for different ways to promote education in distance learning. Furthermore, teachers, instructors, and professors are engaging in ICT integration in teaching the students to protect them from COVID-19.

According to Barbosa (2015), there is an increasing trend in the use of 3D and simulation technologies due to the declining cost of computers as well as the constant increase in processing power. Barbosa (2015) also mentioned that this kind of technology with the right interface and environment can be an asset for the teaching and learning process. Furthermore, simplicity, ease of navigation, and the human-computer interaction aspect of 3D simulation tools allow a student to engage, increase enthusiasm, and better focus towards the learning objective of an activity. This was supported by Azer and Azer (2016) whereas they mentioned in their study that there is an increasing interest in the use of 3D models as evidenced by a large number of publications related to 3D technology. Azer and Azer (2016) further elaborated that 3D models in both traditional and digital formats are favored by most students from healthcare institutions as it can improve their spatial visualization skills.

In the Philippines, TVL (Technical-Vocational-Livelihood) automotive track is one of the tracks in K to 12 that is continuously growing in the student population. Its goal is to prepare their graduates with enough skills to help them generate their income without the choice of pursuing college. TVL Automotive is implemented in the school year 2016-2017. It has 6 strands namely: agricultural fishery, home economics, information and communications technology, industrial arts, applied and specialized subjects in arts and design, and applied and specialized subjects in sports. Although TVL automotive is selected by many students, TVL automotive still has many problems that the schools and the track itself encountered on its implementation, a large portion of this problem is directly attributed to lack of resources.

The uniqueness of the study among the studies previously conducted is that the study intended to develop an application in consideration with the learner's context particularly their needs concerning the Philippine K to 12 educational settings. Thus, the main objective of the study is to develop a learning module using 3D technology to enhance the capability of the school and the teachers in teaching the core competency needed to acquire the TESDA NC II in Service Engine Mechanical System of TVL Automotive students. Specifically, it aims to (1) provide students a platform to simulate and teach mechanical engine assembly, (2), determine the contents of the system, and (3) evaluate the said system using ISO 25010 quality standards.

## LITERATURE REVIEW

Technology brought many advantages especially when the world is facing a global health crisis such as COVID-19. It provides different systems and applications that may help the different sectors of society, businesses, government, and educational institutions. One of the emerging technologies is the use of 3D related applications where Bhavani, Sheshadri, and Unnikrishnan (2010) mentioned the vast potential of such technologies varying from 3D web-based interactive learning application for medical practitioners (Violante & Vezzetti, 2017), 3D Simulation modeling technology for architectural design (Omar, El Messeidy, & Youssef, 2016), and 3D simulation system (Salman & Hussein, 2016). This kind of technology allows the enhancement in the ability of vocational education as well as making it accessible to economically challenge communities and helping the learners in grasping complex skills and concepts.

Huang and Gui (2015) described that developing an application using a 3D environment has a huge economic benefit. It enabled students to put their knowledge into practice in an integrated way such as acquiring transferable competencies, providing new and engaging learning opportunities, and a way to practice safe real-life professional situations (Cela-Ranilla et al., 2014). Such applications also allow asynchronous learning of laboratory experiments at their own home (Salman & Hussein, 2016) and at the same time apply existing theories, explore, and make their own decisions (Koivisto et al., 2017) without being exposed to hazards (Zhu, 2017). Furthermore, Bai et al. (2012) said that using 3D simulation for instruction helps visual learners promote self-paced learning that provides realistic situations. It is also one of the affordable options to engage students in problem-solving and critical reasoning (Bai et al., 2012) which helps them refresh their knowledge in various situations (Violante & Vezzetti, 2017).

Multiple studies showed that computer simulations enhance the traditional instruction from laboratory activities (Hu, Chi, & Chang, 2018; Zhu, 2017; Azer & Azer, 2016; Ma et al., 2014; Paton & Houghton, 2012; Rutten, Van Joolingen, & Van der Veena., 2012; Lateef, 2010) to constructing an effective learning model (Wang & Chen, 2020). Also, in contrast to video, computer simulations give students the ability to process temporal (Vera & Santos, 2018) and spatial dimensions (Hu et al., 2018) that creates an immersive feeling that can blur the lines between reality and the virtual environment (Omar et al., 2016; Rossmann et al., 2013; Cai, Tay, & Ngo, 2012).

While the impact of 3D applications on learning proven to be relatively positive. In one study, Korakakis et al. (2012) confirmed the contribution of the 3D learning environment in the learning process. Another study showed that using 3D applications in educational settings reduces the learning curve among students (Hu et al., 2018). Additionally, the increase in usage of such applications is directly correlated to the academic performance of the students (Hanna, Richards, & Jacobson, 2014). The engagement and interest among students tend to be higher when the 3D learning environment is used (Can & Simsek, 2015; Paton & Houghton, 2012) as most students have a presumption that using 3D learning applications the experiences gap from virtual and reality can be significantly reduced (Can et al, 2015).

## METHODOLOGY

Mixed-method was used in the study, a series of interviews and observations were conducted to understand more of the problems encountered by both teachers and students in one senior high school in Pampanga, Philippines. This public school was selected due to the number of students enrolled in the TVL Automotive track. Interviews with the teachers and assessors are carried out to determine which part of the NC II core competency test of the service engine mechanical system should be added to the application. The response of the teachers and assessors were transcribed and group together to form the main concepts which determine the most needed features of the application, how the application should look like, its core contents, and which specific mechanical parts are to be added in the application.

Towards this goal, the application was developed using C# programming as its base language as well as Unity 3D and SketchUp for the overall interface and 3D models (i.e., machine equipment, engines, and mechanical tools). For the technical specification, the following were used: 64bit Windows 10 operating system, 2.4GHz Intel i7 $5^{th}$ Gen. processor, 8 Gb of RAM, and 2 Gb dedicated Graphics card. While Agile model (i.e., Requirement Analysis, Design, Development, Quality Assurance, and Deployment) was adopted for the development process to ensure the suitability of the learning module. The requirement analysis, design, development, and quality assurance phases were conducted simultaneously before the deployment of the learning module for the alpha testing. The alpha test was administered to determine if the application follows the standard set by the ISO 25010. Three ICT experts and a total of 153 students enrolled in the TVL automotive track evaluated the application using a 5-point Likert scale.

## RESULTS AND DISCUSSION

Based on the interviews, the application should be divided into four (4) main categories namely: Module, Tools, Assemble, and Test category as seen in Figure 1. These sections were highly sought by the informants.

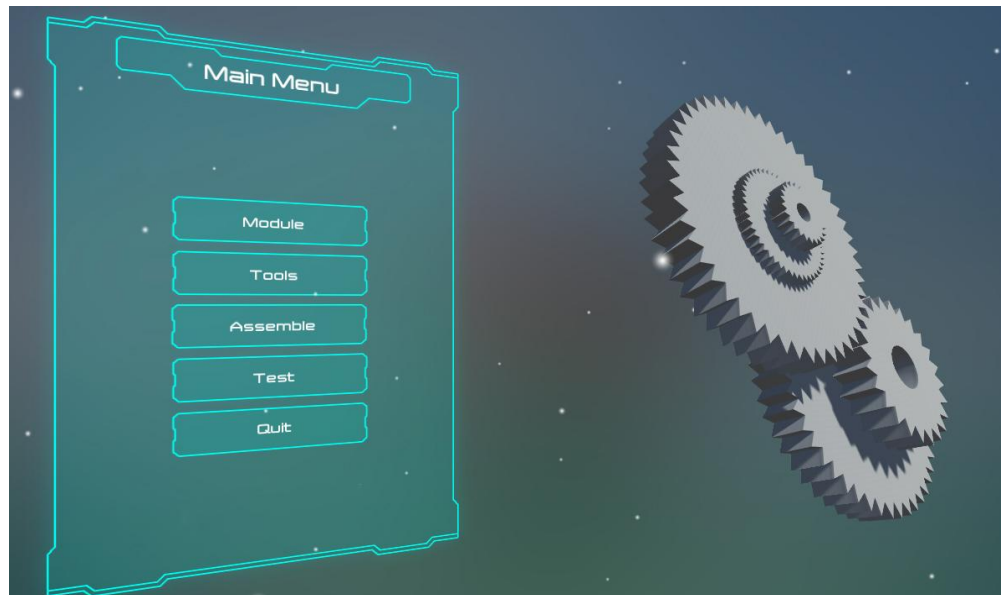

*Figure 1.* The main interface of the learning module after opening the application

The Module category contains the learning materials for engine servicing. Figure 2 shows the category is further subdivided into seven (7) sections, the first part contains all the video demonstrations (i.e., conditions before assembly, disassembly and reassembly procedures, inspections and requirements, and functional check and adjustments) required by the informants to be added in the module. This section is added to ensure that different mediums for learning are present inside the application.

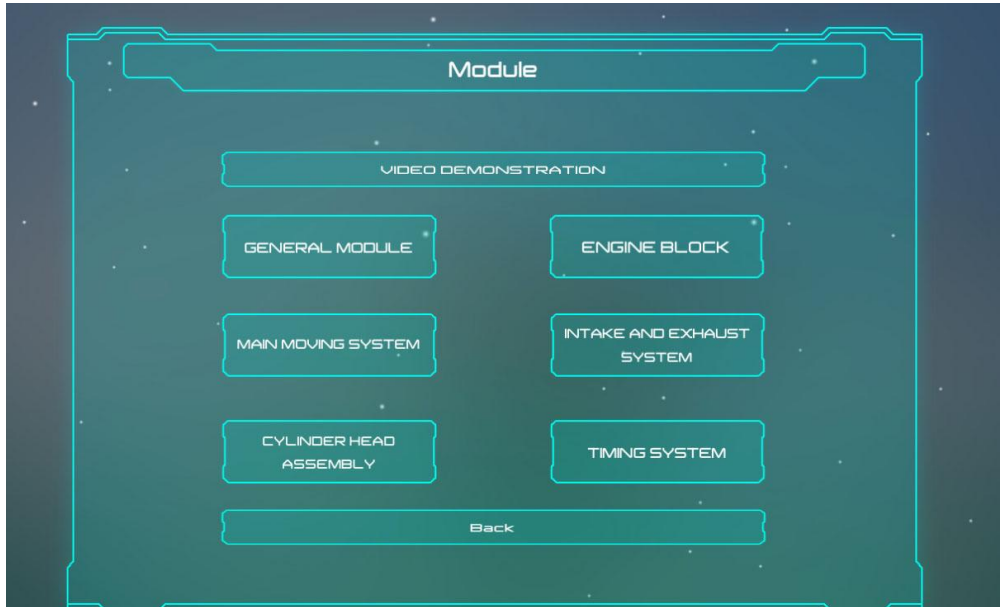

*Figure 2.* Module Section

The general module section contains the brief description and specific specification of varying parts used for engine servicing (e.g., valve spring. Piston, and connecting rod). While the engine block (Figure 3), the main moving system (Figure 4). The intake and exhaust system (Figure 5), cylinder head assembly (Figure 6), and timing system (Figure 7) sections cover the learning materials used by the informants which are approved by the technical education authority.

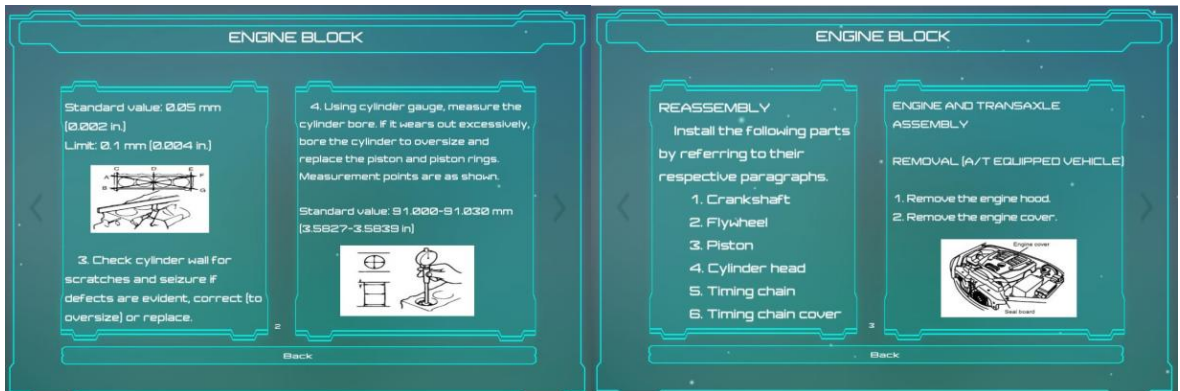

*Figure 3.* Engine Block section interface

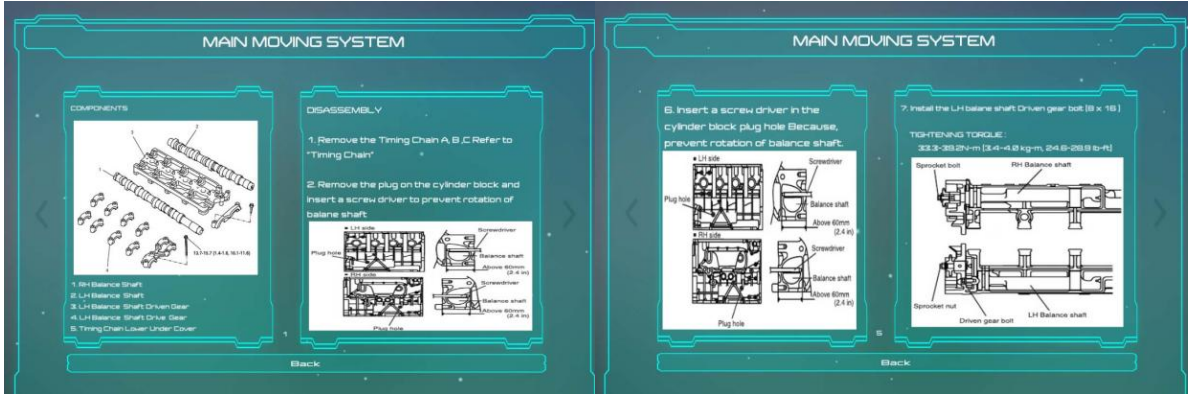

*Figure 4.* Main moving system section interface

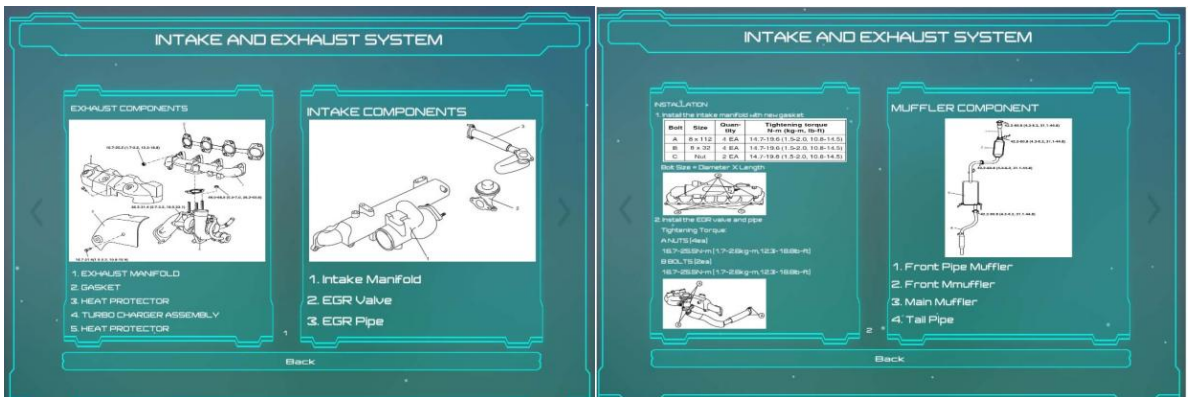

*Figure 5.* Intake and Exhaust section interface

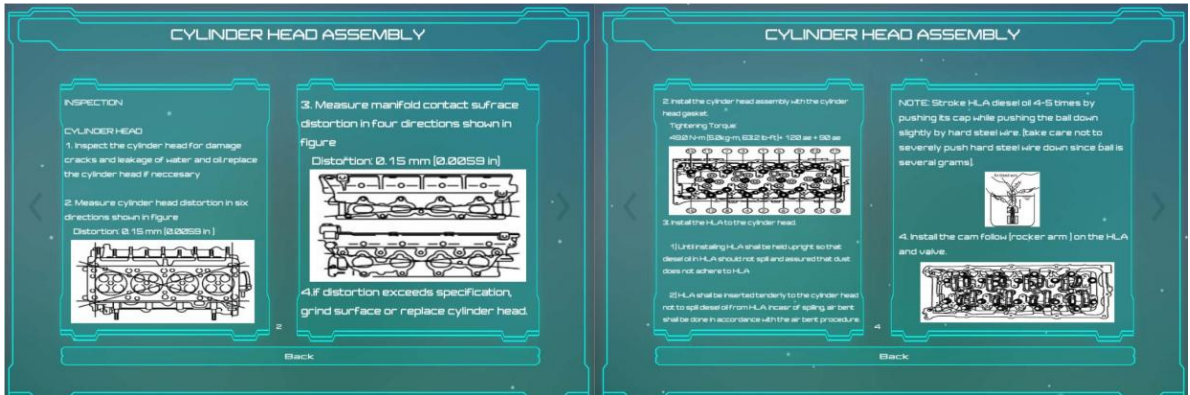

*Figure 6* Cylinder head assembly section interface

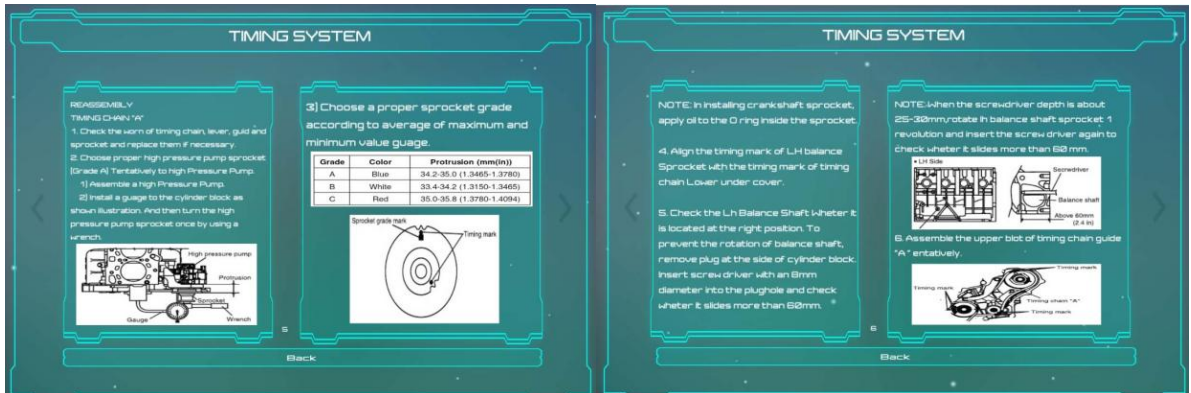

*Figure 7.* Timing system section interface

The Tools category contains the specific tools used in engine servicing, it is subdivided into six (6) sections: personal protective equipment (PPE), lubrication equipment, engine assembly, and disassembly tools, engine repair and repower tools, and the engine reassembly and reinstallation. Each section contains 3D animated models and a brief description as seen in Figure 8; these contents are validated by experts in the field.

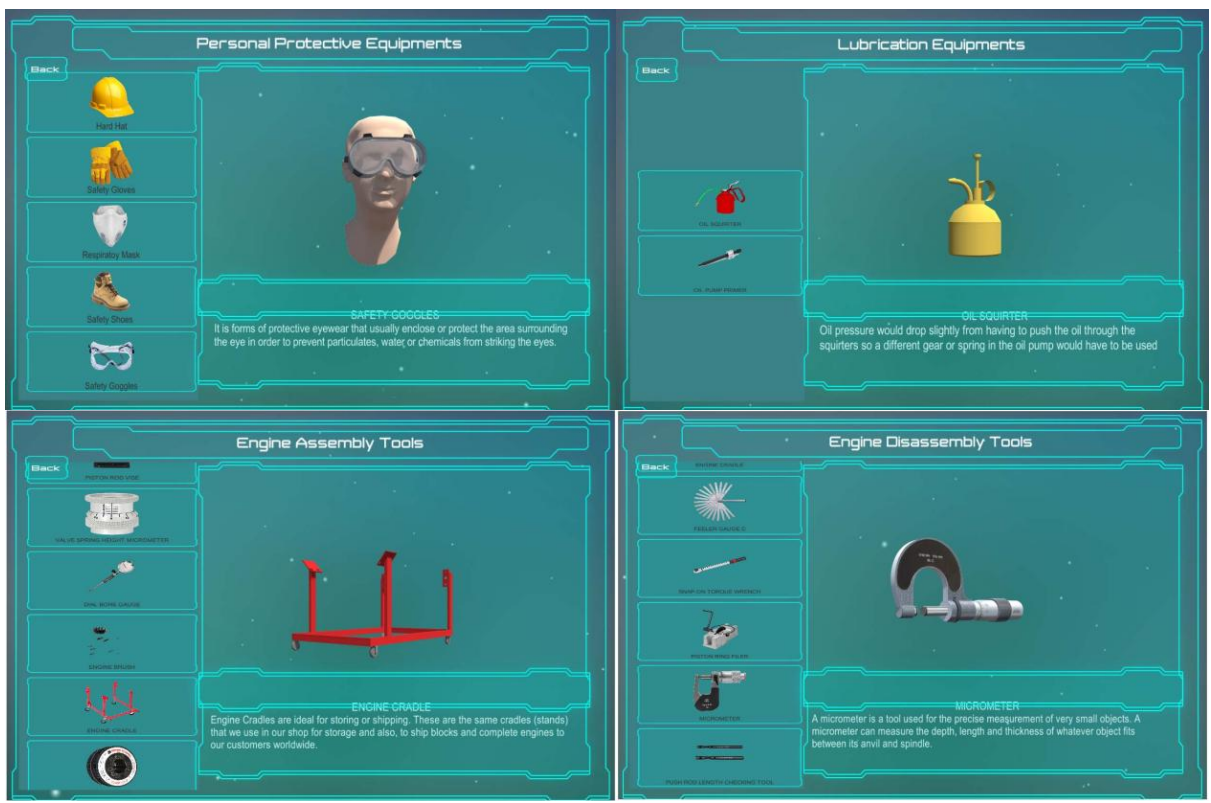

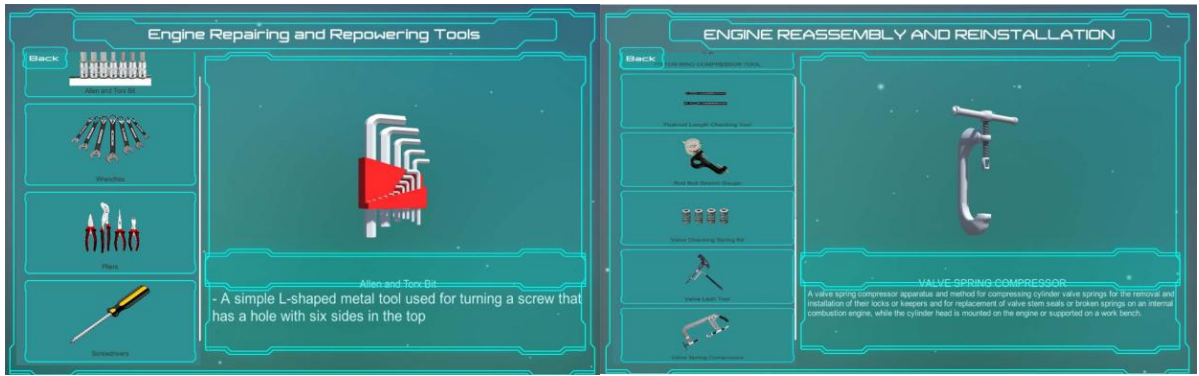
*Figure 8*. The interface of the six sections under the Tools category

Figure 9 shows the interface where learners and teachers can simulate engine assembly. The window has three (3) main parts. From the left side shows the different tools that can be used to assemble the engine, the bottom part shows the steps needed to assemble the engine, and lastly the middle part where the user learners can simulate the engine assembly. Basic mouse and keyboard controls (i.e., Q – rotate to the X-axis, W – rotate to the Y-axis, E – rotate to the Z-axis) could be used to rotate the 3D objects. The learners may also opt to use the numeric keypad to change their perspective (i.e., Front, Back, Top, Bottom, Left, Right).

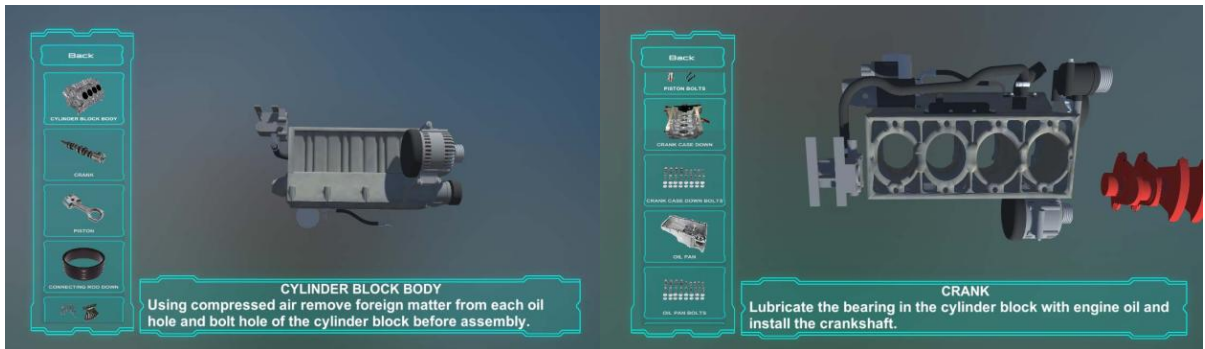
*Figure 9*. The interface of the Assembly category

Figure 10 displayed how the assessment part of the system looks like. It contains randomized content for testing the understanding and familiarity of the students towards the engine parts. It has 4 mains parts namely: Question (left side), Object (right side), Choices (bottom), and Score (top right). The content of the test was validated and checked by the informants and experts before adding them to the test bank.

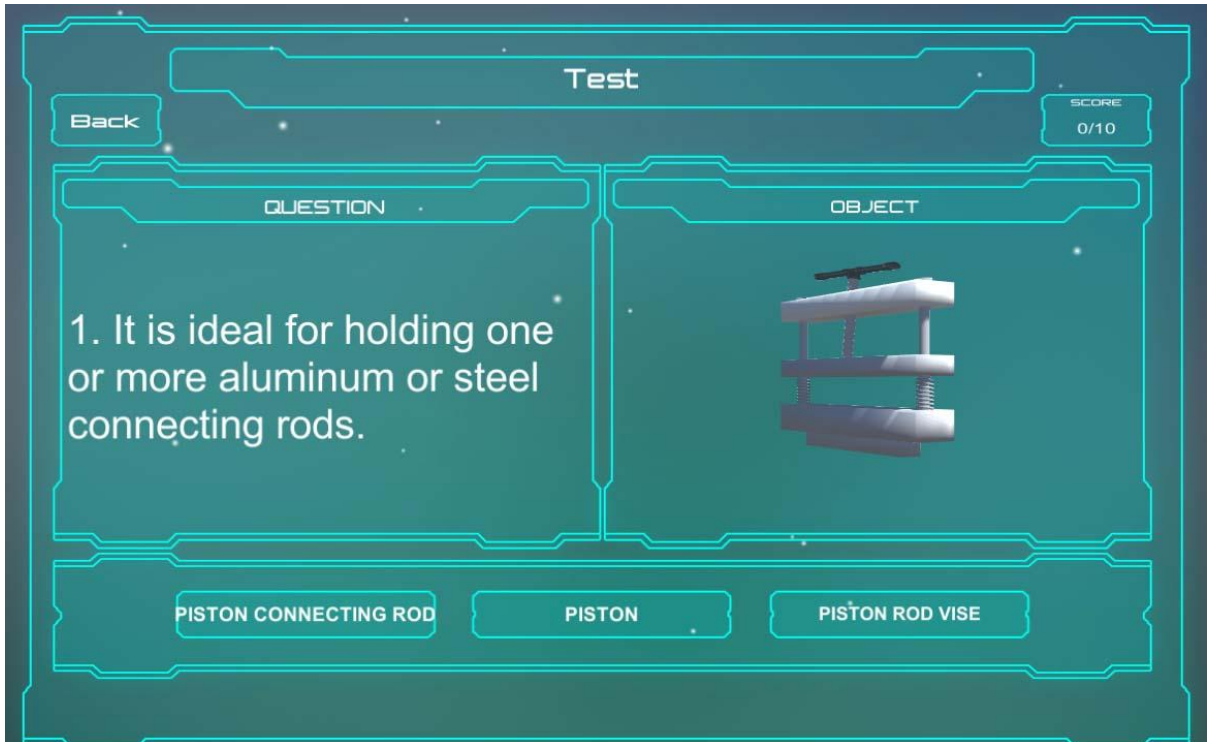

*Figure 10*. The interface of the Test category

Table 1 illustrates that the application is rated as moderately acceptable by its intended users (i.e., students) and by the ICT experts in the alpha testing. It also means that the application follows the ISO 25010 quality standards.

Table 1. ISO 25010 Results of the Alpha Test

| Quality Standards | Mean ($\bar{x}$) | Interpretation |
|---|---|---|
| Functional Suitability | 3.94 | Moderately Acceptable |
| Performance Efficiency | 3.84 | Moderately Acceptable |
| Compatibility | 4.13 | Moderately Acceptable |
| Usability | 3.54 | Moderately Acceptable |
| Reliability | 2.82 | Acceptable |
| Security | 2.56 | Acceptable |
| Maintainability | 3.35 | Acceptable |
| Portability | 4.01 | Acceptable |
| Mean | 3.52 | Moderately Acceptable |

*Legend: 4.51 – 5.0 (Highly Acceptable); 3.51 – 4.50 (Moderately Acceptable); 2.51 – 3.50 (Acceptable); 1.51 – 2.50 (Slightly Acceptable); 1.0 – 1.51 (Not Acceptable)*

## CONCLUSIONS AND RECOMMENDATIONS

Since most of the public schools contain one or more computer laboratories before the implementation of K to 12 curricula, the study underscores the need and applicability of using digital learning modules to enhance the capability of teachers in teaching technical skills and solves the deficiency of equipment and tools which can be used by students. It is also worth mentioning that the students and teachers both agreed that the application is suitable and helpful together with the other learning materials towards learning and understanding their lessons.

For future studies, it is recommended that additional car types and parts could be added to the application given the growing variety of automobiles in the industry as well as making it customizable and a mobile version of the application is also worth pursuing. Also, future researchers can study the effectiveness and impact of the developed application by subjecting it to Usability Acceptance Testing (UAT) in the actual setting. Additionally, they should also look at the potential use of augmented, virtual, and/or mixed reality to the field, particularly in the test/assessment section.

## IMPLICATIONS

The study showed that public education institutions could harness the potential of emerging technologies like 3D as a way to teach students in learning laboratory and practical subjects.

## ACKNOWLEDGEMENT

The researchers are indebted to the informants who help in the completion of this study and to Don Honorio Ventura State University for their support in this endeavor.